\documentclass[conference]{IEEEtran}
\IEEEoverridecommandlockouts
\usepackage{cite}
\usepackage{amsmath,amssymb,amsfonts}
\pdfoutput=1

\usepackage{booktabs}
\usepackage{graphicx}
\usepackage{textcomp}
\usepackage{xcolor}
\usepackage{algorithm}
\usepackage{algpseudocode}

\def\BibTeX{{\rm B\kern-.05em{\sc i\kern-.025em b}\kern-.08em
    T\kern-.1667em\lower.7ex\hbox{E}\kern-.125emX}}
\begin{document}

\title{A Fluid Antenna Enabled Physical Layer Key Generation for Next-G Wireless Networks
}
\author{
		\IEEEauthorblockN{Jiacheng Guo\IEEEauthorrefmark{1},~Ning Gao\IEEEauthorrefmark{1},~Yiping Zuo\IEEEauthorrefmark{2},~Hao Xu\IEEEauthorrefmark{3},~Shi Jin\IEEEauthorrefmark{3}~and Kai Kit Wong\IEEEauthorrefmark{4}}
		
		\IEEEauthorblockA{\IEEEauthorrefmark{1}School of Cyber Science and Engineering, Southeast University, Nanjing 210096, China.
		}
		\IEEEauthorblockA{\IEEEauthorrefmark{2}Nanjing University of Posts and Telecommunications, Nanjing 210023, China.\\
		}
	\IEEEauthorblockA{\IEEEauthorrefmark{3}National Mobile Communications Research Laboratory, Southeast University, Nanjing 210096, China.
		}		
	\IEEEauthorblockA{\IEEEauthorrefmark{4}Department
of Electronic and Electrical Engineering, University College London, WC1E 6BT London, U.K.
		}
		\IEEEauthorblockA{E-mails:\{220245444, ninggao, hao.xu, jinshi\}@seu.edu.cn, zuoyiping@njupt.edu.cn, kai-kit.wong@ucl.ac.uk}
	}
\maketitle

\begin{abstract}
As a promising physical layer security technique, physical layer key generation (PLKG) enables legitimate users to obtain secret keys from wireless channel without security infrastructures. However, in harsh propagation environments, the channel characteristic becomes unsatisfactory, the key generation rate (KGR) is significantly deteriorated. In this paper, we propose a novel fluid antenna (FA) enabled PLKG system to address this challenge. Specifically, we first derive the closed-form expression of the KGR for FA array, and then jointly optimize the precoding matrix and the antenna positions via a particle swarm optimization (PSO) algorithm. Next, to further reduce the computational complexity of the optimization procedure, we develop an alternating optimization (AO) algorithm, which combines the projected gradient descent (PGD) and the PSO. Simulation results demonstrate that by exploiting the additional spatial degree of freedom (DoF), our FA enabled PLKG system is superior to the benchmarks, such as the conventional fixed-position antenna (FPA) array and the reconfigurable intelligent
surface (RIS). It is worth highlighting that compared to the conventional uniform planar antenna (UPA), the FA enabled PLKG achieves a 35.42\% KGR performance
improvement under PSO algorithm and a
67.73\% KGR performance improvement under AO algorithm, respectively.
\end{abstract}


\section{Introduction}
As the fifth-generation (5G) communications become commercially available, the research on sixth-generation (6G) communications is also advancing actively. 6G not only inherits many of the advantages of 5G, but also introduces a series of emerging communication technologies such as the endogenous artificial intelligence (AI) communications, the integrated
sensing and communication (ISAC) and the reconfigurable antenna, etc \cite{SCISGAOllm,10004900,10682524}. However, these emerging communication technologies have increased new attack surfaces for wireless networks, which bring new security and privacy challenges \cite{10044183}. In recent years, the data security has become the cornerstone of the digital circulation era. Due to the broadcast nature of wireless networks, the wireless data transmitted over the air is vulnerable to security threats such as eavesdropping, spoofing and jamming, etc. With the rapid proliferation of wireless devices and the growing demand for instant connectivity, there is an increasing demand for data encrypted communications. However, the traditional data encryption heavily relies on pre-distributed key and trusted public infrastructures, making them less suitable for dynamic and massive real-time communications. As an promising alternative scheme, physical layer key generation (PLKG) leverages the inherent reciprocity and the randomness of wireless propagation environments, enabling two legitimate parties to independently generate shared secret keys. Since the spatial decorrelation of the wireless channels, eavesdropper located beyond half a wavelength cannot obtain the same channel characteristic, i.e., channel state information
(CSI), thereby achieving information theoretic security without relying on the computational complexity.

The basic process of PLKG consists of four main steps: channel probing, quantization, information reconciliation, and privacy amplification \cite{10475842}. During the channel probing, Alice and Bob transmit the pilot signal to share the CSI in a channel coherence time. And then, in the quantization phase, the CSI extracted by Alice and Bob is quantized into the raw bit sequences. Next, during the information reconciliation, the error correction codes are used to correct discrepancies in the raw bit sequences between Alice and Bob. Finally, the privacy amplification is employed to remove bits that can be leaked to eavesdropper, which is to enhance the data security at the physical layer. Although the effectiveness of PLKG has been demonstrated
both theoretically and experimentally, the performance of PLKG can  significantly decline in harsh propagation environments. For example, in an indoor quasi-static wireless environments, negotiating a sufficiently random raw key is time-consuming task due to the fact that the channel based attenuations are almost predictable \cite{10475842, b9,b5,b6}. The randomness of wireless channel is a critical internal condition for PLKG, which influences the key generation rate (KGR). In other words, the randomness of the wireless channel determines the strength of the key's entropy source. In the early works, a single relay or cooperative relay has been proposed to increase the KGR, but these methods are not suitable for the untrusted relays. On the other hand, the artificial random source has also been used to assist the PLKG, but these additional designs can lead to the incompatibility for the commercial communication protocols \cite{6582550}.

Reconfigurable intelligent surface (RIS) is one of the programmable metamaterial antenna, which can customize the wireless channel characteristics, i.e., randomness \cite{10475842}. In recent years, the RIS has been used to improve the performance of PLKG, which can achieve a good balance between trustworthy and compatibility. Reference \cite{b5} has utilized RIS to maximize the total KGR in a multi-user system. A new integrated
communications and security (ICAS) paradigm has been proposed in \cite{b6}, which conducted the first study on the ``one-time pad" communication with the assistance of RIS. However, this manufacturer pre-determined antenna arrangement structure still insufficient to further exploit the potential of PLKG in harsh propagation environments. Very recently, fluid antenna (FA) system have garnered increasing attention in the field of wireless communications due to their ability to dynamically adjust antenna positions to enhance the communication performance  \cite{b3}. The authors in \cite{b4} jointly optimize the antenna position, the offloading ratio, and the CPU allocation frequency for MEC servers, which is to minimize the total system latency for all users in the MEC system. Reference \cite{b7} jointly optimizes the antenna position and the beamforming to maximize the secrecy rate. In terms of PLKG, by adjusting the spatial freedom of FA and the phase-shift, the entropy of PLKG can be changed, and thus the FA can hopefully improve the KGR. However, to the best of the authors' knowledge, this novel application has not been studied before.

To this end, in this paper, we study the unique advantages of FA in improving the KGR under the quasi-static wireless environments. Specifically, we develop the FA enabled PLKG
system and derive the closed-form expression of KGR. And then, we jointly optimize the antenna position and the precoding to improve the KGR based on the proposed particle swarm optimization (PSO) algorithm and alternating optimization (AO) algorithm, respectively. The simulation results show that compared with the benchmarks, the FA enabled PLKG achieves a 35.42\% KGR performance improvement under particle
swarm optimization (PSO) algorithm and a 67.73\% KGR performance improvement under AO algorithm, which demonstrates the huge potential of FA array for PLKG.

\section{System Model}

We consider secure transmission in an FA enabled wireless communication system, which is depicted in Fig. 1. In this system, a base station (Alice) with $N$ FAs sends the  message to a single-antenna legitimate user (Bob) over the wireless channel. As shown in this figure, the FAs are connected to RF chains via the flexible cables, and thus their positions can be adjusted. The positions of the $n$th FA can be defined as $\mathbf{t}_n = [x_n, y_n]^T \in \mathcal{D}$ for $n = 1, \ldots, N$, where $\mathcal{D}$ denotes the two-dimensional removable region of FA. The main notations of this paper can be found in Table \ref{table_para}.

\begin{table}[htbp]
	\centering
	\caption{Main notations}
	\label{tab:notation}
	\begin{tabular}{@{}ll@{}}
		\toprule
		\textbf{Symbol} & \textbf{Description} \\
		\midrule
		$x$ & Scalar \\
		$\mathbf{x}, \mathbf{X}$ & Vector and matrix \\
		$\hat{\{\cdot\}}$ & An estimated quantity  \\
		$\{\cdot\}^*$ & Complex conjugate \\
		$\{\cdot\}^T$ & Matrix transpose \\
		$\{\cdot\}^H$ & Matrix conjugate transpose  \\
		$j = \sqrt{-1}$ & Unit imaginary number \\
		$x_{ij}$ & Vector or matrix element \\
		$\mathbf{I}$ & Identity matrix \\
		$I(X;Y)$ & Mutual information of $X$ and $Y$ \\
		$|\mathbf{X}|$ & Determinant of $\mathbf{X}$ \\
		$|x|$ & Absolute value of $x$ \\
		$\mathbb{E}\{\cdot\}$ & Expectation \\
		$\sum$ & Sum operator \\
		$\mathcal{D}$ & Feasible domain for antenna positions \\
		$\mathrm{Tr}(\cdot)$ & Trace of matrix \\
		$\|\cdot\|$ & Euclidian norm operator \\
		$\mathcal{M}_m$ & Position vector of the $m$-th particle\\
		$\mathcal{V}_m$ & Velocity vector of the $m$-th particle\\
		$\odot$ & Hadamard product\\
        $O(\cdot)$ & Order of computational complexity\\
		\bottomrule
	\end{tabular}
	\label{table_para}
\end{table}

\begin{figure}[htbp]
	\centering
	\includegraphics[width=0.51\textwidth]{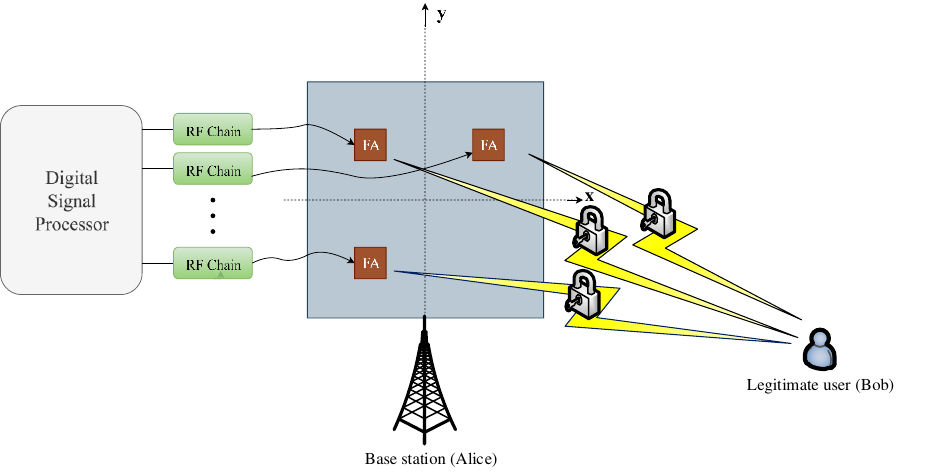} 
	\caption{The schematic diagram of a FA enabled secure communication system.}
	\label{fig:system_model}
\end{figure}
\subsection{Channel Model}\label{AA}
We assume a quasi-static block-fading wireless channel and focus on a specific fading block, where the multipath channel components remain unchanged at a fixed time block.  With the assistance of the FAs, the channel is reconfigurable by adjusting the positions of FAs. Define the set of the coordinates of $N$ FAs by $\mathbf{T} = [\mathbf{t}_1 \ldots \mathbf{t}_N] \in \mathbb{R}^{2 \times N}$. Then, the Alice-to-Bob channel vector is given by $\mathbf{h}_{AB}(\mathbf{T}) \in \mathbb{C}^{N \times 1}$, which are the functions of $\mathbf{T}$. It is worth noting that Alice-to-Bob channel vectors are determined by the signal propagation environments and the positions of FAs. The channel response of FAs can be given by
\begin{equation}
\mathbf{h}_{AB}(\mathbf{T}) = [h(\mathbf{t}_1) \ldots h(\mathbf{t}_N)]^T,
\label{eq:h_T}
\end{equation}
where $
h(\mathbf{t}_n) = \sum_{l=1}^{L} \sigma_l e^{j \frac{2\pi}{\lambda}\mathbf{t}_n^T \boldsymbol{\rho}_{l}}$,
$\boldsymbol{\rho}_{l} = [\sin \theta_{l} \cos \phi_{l}, \cos \theta_{l}]^T$ is the unit
direction vector of the $l$th propagation path, $L$ is the number of propagation paths, $\theta_{l} \in [0, \pi]$ and $\phi_{l} \in [0, \pi]$ are the elevation and azimuth angles of the $l$th path, respectively, $\sigma_l$ is the associated path gain, and $\lambda$ is the wavelength.

The channel probing contains two phases, i.e., the downlink phase and the uplink phase. According to the channel reciprocity, we can assume that $\mathbf{h}_{\mathrm{AB}}^T=\mathbf{h}_{\mathrm{BA}}$. Then, we consider the uplink channel probing from Bob to Alice, where the received signal at Alice can be expressed as
\begin{equation}
	\mathbf{y}_{BA} = \mathbf{h}_{BA}^T s_u + \mathbf{n}_a = \mathbf{h}_{AB} s_u + \mathbf{n}_a, \quad
\end{equation}
where $\mathbf{y}_{BA} \in \mathbb{C}^{N \times 1}$, $s_u$ is the transmission signal of Bob, $\mathbf{n}_a$ is the complex Gaussian noise vector with each element zero mean and variance $\sigma^2$.

During the uplink channel probing phase, Alice obtains the estimated CSI via the least squares (LS) channel estimation, which is denoted as
\begin{equation}
	\bar{\mathbf{y}}_{BA} = \mathbf{h}_{AB} + \mathbf{n}_a s_u^*.
\end{equation}
We assume that the pilot length is $S$. Then, Alice multiplies the channel estimation with the precoding matrix $
	\mathbf{P} \in \mathbb{C}^{N \times S}$
 and thus receives the signal $\hat{\mathbf{y}}_{BA}$, which can be written as
\begin{equation}
	\hat{\mathbf{y}}_{BA} = \mathbf{P}^T \mathbf{h}_{AB} + \mathbf{P}^T \mathbf{n}_a s_u^*.
\end{equation}
In the downlink phase, Alice transmits a pilot signal of length $N$, $\mathbf{S}_d \in \mathbb{C}^{N \times N}$ with $\mathbf{S}_d^H \mathbf{S}_d = \mathbf{I}_N$ to Bob. The received signal at Bob is given by \cite{b8}
\begin{equation}
	\mathbf{y}_{AB} = \mathbf{h}_{AB}^T \mathbf{P} \mathbf{S}_d^H + \mathbf{n}_b,
\end{equation}
where $\mathbf{y}_{AB} \in \mathbb{C}^{1 \times S}$, $\mathbf{n}_b \in \mathbb{C}^{1 \times N}$ is the complex Gaussian noise vector at Bob with each element zero mean and variance $\sigma^2$. After the LS channel estimation, the downlink channel at Bob is estimated as
\begin{equation}
\hat{\mathbf{y}}_{AB} = \mathbf{y}_{AB} \mathbf{S}_d \left( \mathbf{S}_d^H \mathbf{S}_d \right)^{-1}
= \mathbf{h}_{AB}^T \mathbf{P} + \mathbf{n}_b \mathbf{S}_d.
\end{equation}

\subsection{ Problem Formulation}
The optimization problem can be formulated as follows. We aim to maximize the secret key rate by jointly optimizing the BS's precoding matrix $\mathbf{P}$ and the antennas' position $\mathbf{T}$. To this end, the optimization problem is formulated as
\begin{equation}
	\begin{aligned}
		\mathcal{P}: \max_{\mathbf{P}, \mathbf{T}} \quad & R_{\mathrm{sk}}, \\
		\text{s.t.} \quad & \|\mathbf{t}_i - \mathbf{t}_j\| \geq d_{\min}, \forall~1 \leq i \neq j \leq N, \\
		& \mathbf{t}_n \in \mathcal{D}, \\
		& \mathrm{Tr}(\mathbf{P} \mathbf{P}^{H})= P_{\max},
	\end{aligned}
\end{equation}
where $R_{\mathrm{sk}} $ denotes KGR  and $P_{\max}$ is the total transmit power constraint of Alice, a minimum distance $d_{\min}$ is required to avoid the coupling effect between each pair of antennas in the transmit region.

 \section{Proposed Solutions}
 \subsection{ Secret Key Rate for FA}
KGR is defined as the maximum number of secret key
bits that can be extracted from a wireless channel probing \cite{b9}. The exact expression of secret key generation is still an open problem, but can be approximately calculated by the mutual information \cite{5483148}. Due to the loss of bit quantization, we have the mutual information $I(\mathbf{\hat{y}}_{AB}; \mathbf{\hat{y}}_{BA}) \geq I(\mathbf{q}_A; \mathbf{q}_B)$. Given the channel reciprocity of all links in time division duplex (TDD) systems, the KGR from Alice to Bob, and Bob to Alice are equal. Therefore, we assume that Bob is the active initiator of key generation, and thus, we take the key generation rate at Alice side as a study case. In this case, the KGR can be characterized as
\begin{equation}
\label{RSK}
R_{\mathrm{sk}} = I(\hat{\mathbf{y}}_{\mathrm{BA}}; \hat{\mathbf{y}}_{\mathrm{AB}}^T)
= \log_2 \left( \frac{|\mathbf{R}_a||\mathbf{R}_b|}{|\mathfrak{R}_{a,b}|} \right),
\end{equation}
where
\begin{align*}
	&\mathbf{R}_a = \mathbb{E} \left\{ \hat{\mathbf{y}}_{\mathrm{BA}} \hat{\mathbf{y}}_{\mathrm{BA}}^{H} \right\} = \mathbf{P}^{T} \mathbb{E} \left\{ \mathbf{h}_{\mathrm{AB}} \mathbf{h}_{\mathrm{AB}}^{H} \right\} \mathbf{P}^{*} + \sigma^2 \mathbf{P}^{T} \mathbf{P}^{*}, \\
	&\mathbf{R}_b = \mathbb{E} \left\{ \hat{\mathbf{y}}_{\mathrm{AB}}^{T} \hat{\mathbf{y}}_{\mathrm{AB}}^{*} \right\} = \mathbf{P}^{T} \mathbb{E} \left\{ \mathbf{h}_{\mathrm{AB}} \mathbf{h}_{\mathrm{AB}}^{H} \right\} \mathbf{P}^{*} + \sigma^2 \mathbf{I}_N, \\
	&\mathfrak{R}_{a,b} = \begin{bmatrix}
		\mathbf{R}_a & \mathbf{R}_{a,b} \\
		\mathbf{R}_{b,a} & \mathbf{R}_b
	\end{bmatrix}.
\end{align*}
Note that $\mathbf{R}_{a,b} = \mathbf{R}_{b,a}=\mathbf{R}$, we have
\begin{align*}
\mathbf{R}
= \mathbb{E} \left\{ \hat{\mathbf{y}}_{\mathrm{BA}} \hat{\mathbf{y}}_{\mathrm{AB}}^{*} \right\}
= \mathbf{P}^{T} \mathbb{E} \left\{ \mathbf{h}_{\mathrm{AB}} \mathbf{h}_{\mathrm{AB}}^{H} \right\} \mathbf{P}^{*},
\end{align*}
where each element in the matrix $\mathbf{R}$ can be denoted as
 \begin{equation}
 \mathbf{R}[i,j] = \mathbb{E} \left[ \left( \sum_{l=1}^{L} \sigma_l e^{j \frac{2\pi}{\lambda} \mathbf{t}_i^{T} \boldsymbol{\rho}_l} \right) \left( \sum_{m=1}^{L} \sigma_m^{*} e^{-j \frac{2\pi}{\lambda} \mathbf{t}_j^{T} \boldsymbol{\rho}_m} \right) \right].
 \end{equation}

\subsection{PSO algorithm for problem $\mathcal{P}$}
Due to the high coupling between the position matrix $\mathbf{T}$ and the precoding matrix $\mathbf{P}$, the problem $\mathcal{P}$ is a non-convex optimization problem, the traditional convex optimization methods cannot be directly applied.

The PSO algorithm is a heuristic algorithm derived from research imitating birds' foraging behaviour \cite{b10}. Its basic idea behind this algorithm is to use the sharing of information among individuals in a group to find the optimal solution. Therefore, a PSO-based method can be used to solve optimization problem $\mathcal{P}$. Specifically, we initialize the particle swarm, where each particle represents a combination of optimisation variables, including the real part and the imaginary part of the precoding matrix and the antenna position, which has total $2NS$ variables and $2N$ variables, respectively. The initial positions and velocities of these particles are randomly generated within the specified range. The fitness function of particles is determined by Eq. \eqref{RSK}. By introducing a penalty term to satisfy the minimum distance constraint between FAs,
the final optimization goal of the $m$th particle is
\begin{equation}
f\left(\mathcal{M}_m^{(i)}\right) = R_{\mathrm{sk}}\left(\mathcal{M}_m^{(i)}\right) - \lambda  \left| \mathcal{P} \left( \mathcal{M}_m^{(i)} \right) \right|,
\end{equation}
where $i$ denotes the iteration number, $\lambda$ denotes the penalty coefficient, which is used to ensure minimum distance constraint. The symbol $\mathcal{P} \left( \mathcal{M} \right)$ denotes the antenna spacing constraint penalty function, which is defined as
\begin{equation}
\mathcal{P} (\mathcal{M})  = \sum_{i=1}^{N} \sum_{j=i+1}^{N} \max \left( d_{\min} - \left\| \mathbf{t}_i - \mathbf{t}_j \right\|, 0 \right)^2.
\end{equation}
\begin{algorithm}[htbp]
	\caption{PSO Algorithm for maximizing $R_{\mathrm{sk}}$}
	\label{alg:joint_pso}
	\begin{algorithmic}[1]
		\Require Number of antennas $N$, number of pilots $S$, power constraint $P_{\max}$, noise variance $\sigma^2$, wavelength $\lambda$, number of particles $M$, max iterations $T_{\max}$, PSO parameters $c_1, c_2, \omega_{\max}, \omega_{\min}$;
		\Ensure $\mathbf{P}_{\mathrm{best}}$, $\mathbf{T}_{\mathrm{best}}$, and $R_{\mathrm{sk}}^{\mathrm{best}}$;
		
		\State Initialize particle positions $\mathcal{M}_m^{(0)}$ and velocities $\mathcal{V}_m^{(0)}$ for $m = 1, \dots, M$;
		\For{$m = 1$ \textbf{to} $M$}
		\State Decode $\mathcal{M}_m^{(0)}$ into $\mathbf{P}_m^{(0)}$ and $\mathbf{T}_m^{(0)}$;
		\State Optimize path directions $\{\theta_l, \phi_l\}_{l=1}^L$ from $\mathbf{T}_m^{(0)}$ and construct channel covariance matrix $\mathbf{R}^{(0)}$;
		\State Evaluate fitness $R_m^{(0)} = f(\mathcal{M}_m^{(0)})$;
		\State Set personal best: $\mathcal{M}_{m,\mathrm{best}} = \mathcal{M}_m^{(0)}$, $R_{m,\mathrm{best}} = R_m^{(0)}$;
		\If{$R_m^{(0)} > R_{\mathrm{sk}}^{\mathrm{best}}$}
		\State Update global best: $\mathcal{M}_{\mathrm{best}} = \mathcal{M}_m^{(0)}$, $R_{\mathrm{sk}}^{\mathrm{best}} = R_m^{(0)}$;
		\EndIf
		\EndFor
		
		\For{$i = 1$ \textbf{to} $T_{\max}$}
		\State Update inertia weight: $\omega^{(i)} = \omega_{\max} - \frac{i}{T_{\max}}(\omega_{\max} - \omega_{\min})$;
		\For{$m = 1$ \textbf{to} $M$}
		\State Update velocity:
		\Statex $\mathcal{V}_m^{(i)} \gets \omega^{(i)} \mathcal{V}_m^{(i-1)} + c_1 \mathbf{r}_1 \odot (\mathcal{M}_{m,\mathrm{best}} - \mathcal{M}_m^{(i-1)}) + c_2 \mathbf{r}_2 \odot (\mathcal{M}_{\mathrm{best}} - \mathcal{M}_m^{(i-1)})$;
		\State Update position: $\mathcal{M}_m^{(i)} \gets \mathcal{M}_m^{(i-1)} + \mathcal{V}_m^{(i)}$;
		\State Normalize the precoding matrix $\mathbf{P}_m^{(i)}$ to meet the power constraint;
		\State Optimize path directions and construct channel covariance matrix $\mathbf{R}^{(i)}$;
		\State Evaluate fitness: $R_m^{(i)} = f(\mathcal{M}_m^{(i)})$;
		\If{$R_m^{(i)} > R_{m,\mathrm{best}}$}
		\State Update personal best: $\mathcal{M}_{m,\mathrm{best}} = \mathcal{M}_m^{(i)}$, $R_{m,\mathrm{best}} = R_m^{(i)}$;
		\EndIf
		\If{$R_m^{(i)} >R_{\mathrm{sk}}^{\mathrm{best}}$}
		\State Update global best: $\mathcal{M}_{\mathrm{best}} = \mathcal{M}_m^{(i)}$, $R_{\mathrm{sk}}^{\mathrm{best}} = R_m^{(i)}$;
		\EndIf
		\EndFor
		\EndFor
		
		\State Decode $\mathcal{M}_{\mathrm{best}}$ into $\mathbf{P}_{\mathrm{best}}, \mathbf{T}_{\mathrm{best}}$;
		\State \Return $\mathbf{P}_{\mathrm{best}}, \mathbf{T}_{\mathrm{best}}, R_{\mathrm{sk}}^{\mathrm{best}}$.
	\end{algorithmic}
\end{algorithm}
We adopt the normalized scaling method to ensure the power constraint. Specifically, in each round of iteration, the power of each particle is adjusted by
\begin{equation}
\mathbf{P} \leftarrow \mathbf{P} \sqrt{\frac{P_{\mathrm{max}}}{\text{Tr}(\mathbf{P} \mathbf{P}^H)}}.
\end{equation}
Simultaneously improve the global search capability of the PSO algorithm in the early stage and the local convergence capability in the later stage can increase the overall convergence efficiency. Therefore, we adopt a dynamic inertia weight strategy, which is often used for high-dimensional mixed continuous optimization problems \cite{b11}. Specifically, we utilize a linear decreasing dynamic adjustment strategy, which is given by
\begin{equation}
	w(t) = w_{\max} - \left( \frac{w_{\max} - w_{\min}}{T_{\max}} \right) t,
\end{equation}
where $t$ denotes the number of iterations, $T_{\max}$ denotes the maximum number of iterations, $w(t)$ reflects the linear adjustment strategy, $w_{\max}$ and $ w_{\min}$ are the maximum
and minimum values of $w$, respectively. The overall PSO-based algorithm is summarized in \textbf{Algorithm 1}.

\subsection{AO algorithm for problem $\mathcal{P}$}

We assume that the number of Monte Carlo samples is $M$ and the number of particle is $P_{\mathrm{num}}$. Then, with the maximum number of iterations $T_{\max}$, the computational complexity of the PSO algorithm for maximizing $R_{\mathrm{sk}}$ is approximate to
\begin{equation}
\mathcal{O}(T_{\max}P_{\mathrm{num}}(M L N^2 + N S^2 + S^3)).
\end{equation}
Since Monte Carlo method requires multiple sampling and the spatial dimension significantly increases with the substantial increase in the number of antennas, the PSO algorithm shows a high computational complexity in this situation. Therefore, we next consider an AO algorithm to reduce the computational overhead.

When using the AO algorithm, we first decompose the optimization problem into two subproblems, which are optimization subproblem $\mathcal{P}1$ and subproblem $\mathcal{P}2$. In the subproblem $\mathcal{P}1$, we use the projected gradient descent (PGD) optimization strategy. When the antenna position $\mathbf{T}$ is fixed, the channel covariance matrix $\mathbf{R}(\mathbf{T})$ can be treated as a constant. In this case, the KGR becomes a complex-valued function of the precoding matrix $\mathbf{P}$. To ensure differentiability, we adopt the negative secret key rate as the loss function, and the optimization problem is formulated as follows
\begin{equation}
	\begin{aligned}
	\mathcal{P}1: \min_{\mathbf{P}} \quad \mathcal{L}(\mathbf{P}) = - R_{\mathrm{sk}}(\mathbf{P}), \\
	 \text{s.t.} \quad \mathrm{Tr}(\mathbf{P} \mathbf{P}^H) = P_{\max}.
\end{aligned}
\end{equation}
We split $\mathbf{P}$ into real and imaginary parts, which is given by
\begin{equation}
	\mathbf{P}_{\text{vec}} =
	\begin{bmatrix}
		\mathrm{Re}(\mathbf{P}) \\
		\mathrm{Im}(\mathbf{P})
	\end{bmatrix}
	\in \mathbb{R}^{2NS}.
\end{equation}
After each update round, PGD project the updated precoded matrix back to the power feasible region based on Eq. (13).

In the second subproblem $\mathcal{P}2$, we fix the precoding matrix $\mathbf{P}$ and optimize the antenna
coordinate matrix $\mathbf{T}$. When the precoding matrix $\mathbf{P}$ is fixed, the channel covariance matrix $\mathbf{R}(\mathbf{T})$ still depends on the antenna coordinate matrix $\mathbf{T} \in \mathbb{R}^{N \times 2}$, thereby affecting the KGR. Then, we formulate the following optimization problem
\begin{equation}
	\begin{aligned}
		\mathcal{P}2:\max_{\mathbf{T}} \quad & R_{\mathrm{sk}}(\mathbf{P}, \mathbf{T}), \\
		\text{s.t.} \quad & \|\mathbf{t}_i - \mathbf{t}_j\| \geq d_{\min}, \forall~1 \leq i \neq j \leq N,\\
& \mathbf{t}_n \in \mathcal{D}.
	\end{aligned}
\end{equation}
This problem is non-convex and non-smooth with respect to $\mathbf{T}$, which makes it difficult to solve by using the gradient-based methods. Therefore, we use the PSO algorithm developed in Section III-B, where each particle represents a flattened antenna layout $\mathbf{t} \in \mathbb{R}^{2N}$, and the fitness function can refer to Eq. (11). The complete algorithm process can be found in \textbf{Algorithm 2}.

To analyze the computational complexity, we assume that the number of PGD steps is $T_1$,
and the PSO algorithm uses $P_{\mathrm{num}}$ particles and needs $T_2$ iterations.
Then, the computational complexity of the AO algorithm is calculated by
\begin{equation}
\mathcal{O}\left(\left[T_1 \left(N S^2 + S^3\right)+ T_2 P_{\mathrm{num}}\left(M L N^2 + N S^2 + S^3\right)\right]\right).
\end{equation}
Owing to the faster convergence of PGD, $T_1$ is very small, thus, $T_1+T_2\ll T_{\max}$, and the reduction of the particle number, i.e., $P_{\mathrm{num}}$, in the subproblem $\mathcal{P}2$, making the AO algorithm achieve a lower overall computational complexity.

\begin{algorithm}[htbp]
	\caption{ AO for maximizing $R_{\mathrm{sk}}$}
	\label{alg:ao_single}
	\begin{algorithmic}[1]
		\Require Antenna number $N$, pilot number $S$, power constraint $P_{\max}$, noise variance $\sigma^2$, wavelength $\lambda$, PGD steps $T_1$, PSO iterations $T_2$, PSO parameters $M, c_1, c_2, \omega_{\max}, \omega_{\min}$;
		\Ensure $\mathbf{P}_{\mathrm{best}}$, $\mathbf{T}_{\mathrm{best}}$, and $R_{\mathrm{sk}}^{\mathrm{best}}$;
		
		\State \textbf{Step 1: PGD based Precoding Optimization (Fix $\mathbf{T}$)};
		\State Initialize antenna layout $\mathbf{T}_0$;
		\State Construct channel covariance matrix $\mathbf{R}$ from $\mathbf{T}_0$;
		\State Initialize random $\mathbf{P}^{(0)}$ and project to satisfy $\operatorname{Tr}(\mathbf{P}\mathbf{P}^H) = P_{\max}$;
		\For{$t = 1$ to $T_1$}
		\State Compute gradient of $-R_{\mathrm{sk}}(\mathbf{P}, \mathbf{T}_0)$;
		\State Gradient update and project onto power constraint;
		\EndFor
		\State Set $\mathbf{P}_{\mathrm{best}} \gets \mathbf{P}$;
		
		\State \textbf{Step 2: PSO based Antenna Optimization (Fix $\mathbf{P}_{\mathrm{best}}$)}
		\State Initialize particles position $\mathcal{M}_m^{(0)}$ and velocities $\mathcal{V}_m^{(0)}$ for $m = 1, \dots, M$;
		\For{$m = 1$ to $M$}
		\State Evaluate $R_m^{(0)} = f(\mathbf{P}_{\mathrm{best}}, \mathcal{M}_m^{(0)})$;
		\State Set personal best $\mathcal{M}_{m,\mathrm{best}} = \mathcal{M}_m^{(0)}$;
		\State Update global best if $R_m^{(0)}$ is the highest;
		\EndFor
		
		\For{$i = 1$ to $T_2$}
		\State $\omega^{(i)} = \omega_{\max} - \frac{i}{T_2}(\omega_{\max} - \omega_{\min})$;
		\For{$m = 1$ to $M$}
		\State Update $\mathcal{V}_m^{(i)}$ using personal/global best and random vectors $\mathbf{r}_1, \mathbf{r}_2$;
		\State Update $\mathcal{M}_m^{(i)} = \mathcal{M}_m^{(i-1)} + \mathcal{V}_m^{(i)}$ and enforce spacing constraint;
		\State Evaluate $R_m^{(i)} = f(\mathbf{P}_{\mathrm{best}}, \mathcal{M}_m^{(i)})$;
		\State Update global best if the value improved;
		\EndFor
		\EndFor
		\State Set $\mathbf{T}_{\mathrm{best}} \gets \mathcal{M}_{\mathrm{best}}$;
		
		\State \Return $\mathbf{P}_{\mathrm{best}}, \mathbf{T}_{\mathrm{best}}, R_{\mathrm{sk}}^{\mathrm{best}}$.
	\end{algorithmic}
\end{algorithm}
\section{Simulations}
In this section, simulation results are provided to validate the
effectiveness of the proposed algorithms. The number of BS antennas is set to be $N = 4$, and the number of pilot signals to be $S = 4$. The maximum transmit power is normalized to $P_{\max} = 1$, and the noise power is set to be $\sigma^2 = 0.1$. The minimum spacing between any two antennas is constrained as $d_{\min} = \lambda / 2$. The FA array is optimized within the square region of size $[0, 20\lambda]\times [0, 20\lambda]$, and the number of multipath components is set to be $L = 8$.

\begin{figure}[htbp]
	\centering
	\includegraphics[height=5.5cm]{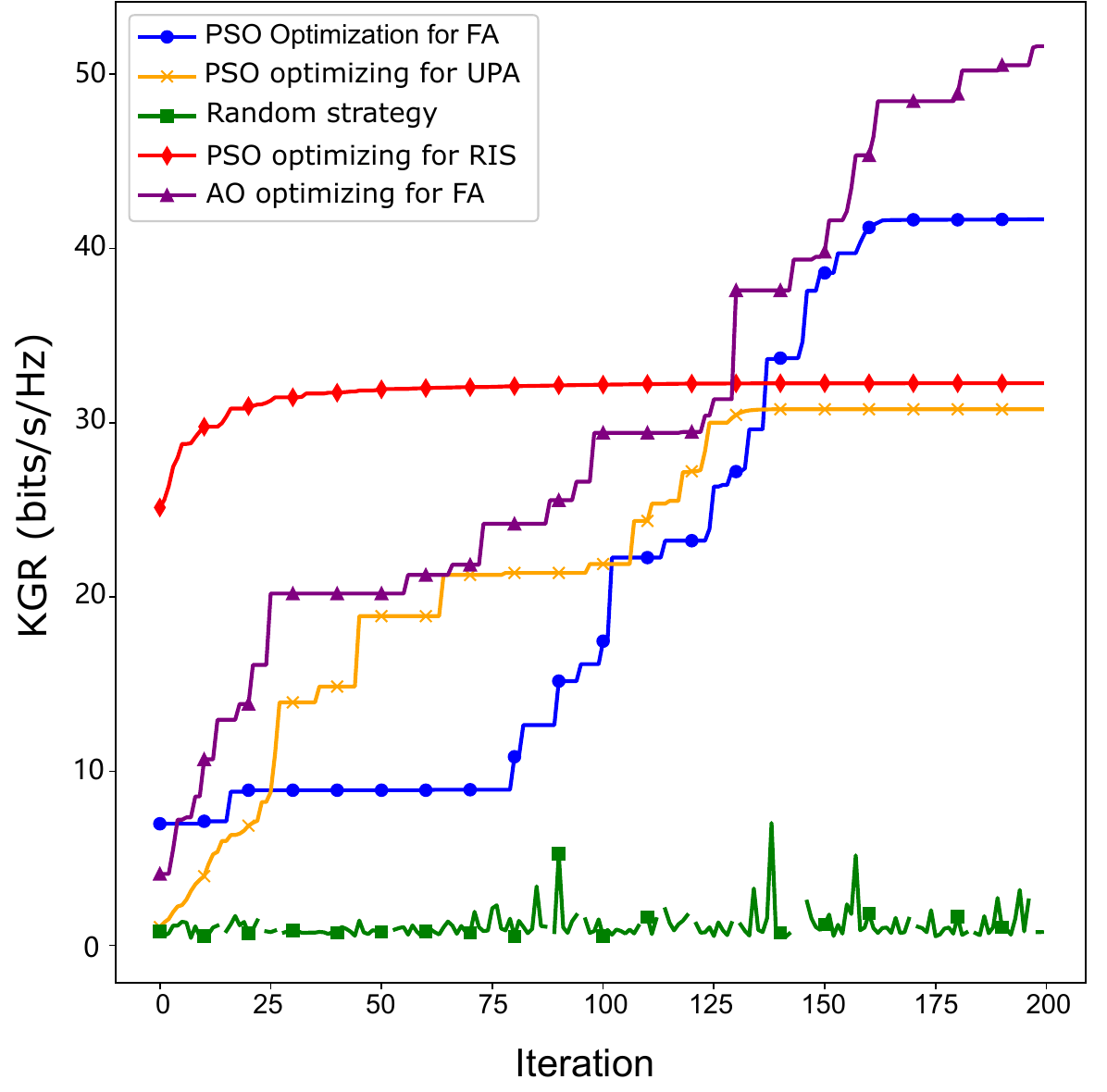} 
	\caption{KGR performance comparison of different methods.}
	\label{fig:comparison}
\end{figure}

For the benchmarks, a random strategy is established by using the independent trials, where both the precoding matrix and antenna positions are randomly generated which satisfies the spacing constraint and power constraint. In addition, to evaluate the effect of the precoding for KGR with the fixed antenna placement, we consider a UPA array as the conventional fixed-position antenna (FPA) scheme, which uses the PSO algorithm presented in \textbf{Algorithm 1}. As shown in reference \cite{10475842}, RIS has shown a tremendous potential for PLKG in harsh propagation environments, we also adopt RIS based PLKG as a benchmark. To ensure the fairness, RIS based PLKG conducts the joint optimization of the precoding matrix and RIS phase shifts with the UPA array.

Figure \ref{fig:comparison} illuminates the KGR of different benchmarks. From this figure, we can observe that the FA based PLKG with AO algorithm achieves the highest KGR. Therein, AO algorithm consistently improves the KGR and converges to the highest value of approximately 52 bits/s/Hz, which demonstrates the effectiveness of the stage-wise variable separation optimization. The PSO algorithm also achieves a good performance where the convergent KGR is slightly lower, i.e., around 42 bits/s/Hz. This is because the high dimensionality of the optimization space causes the PSO algorithm to fall into a local optimum. We can find that RIS based PLKG with PSO algorithm converges to 33 bits/s/Hz with the RIS phase-shift and UPA precoding optimization. The UPA applies PSO algorithm to optimize the precoding matrix, where its performance converges early and the KGR approximately approaches to 31 bits/s/Hz. Lastly, the random strategy randomly samples both precoding and antenna position under the power and spacing constraints. It exhibits highly unstable and the poorest performance, which highlights the necessity of the guided optimization strategies.
\begin{figure}[htbp]
	\centering
	\includegraphics[height=5.5cm]{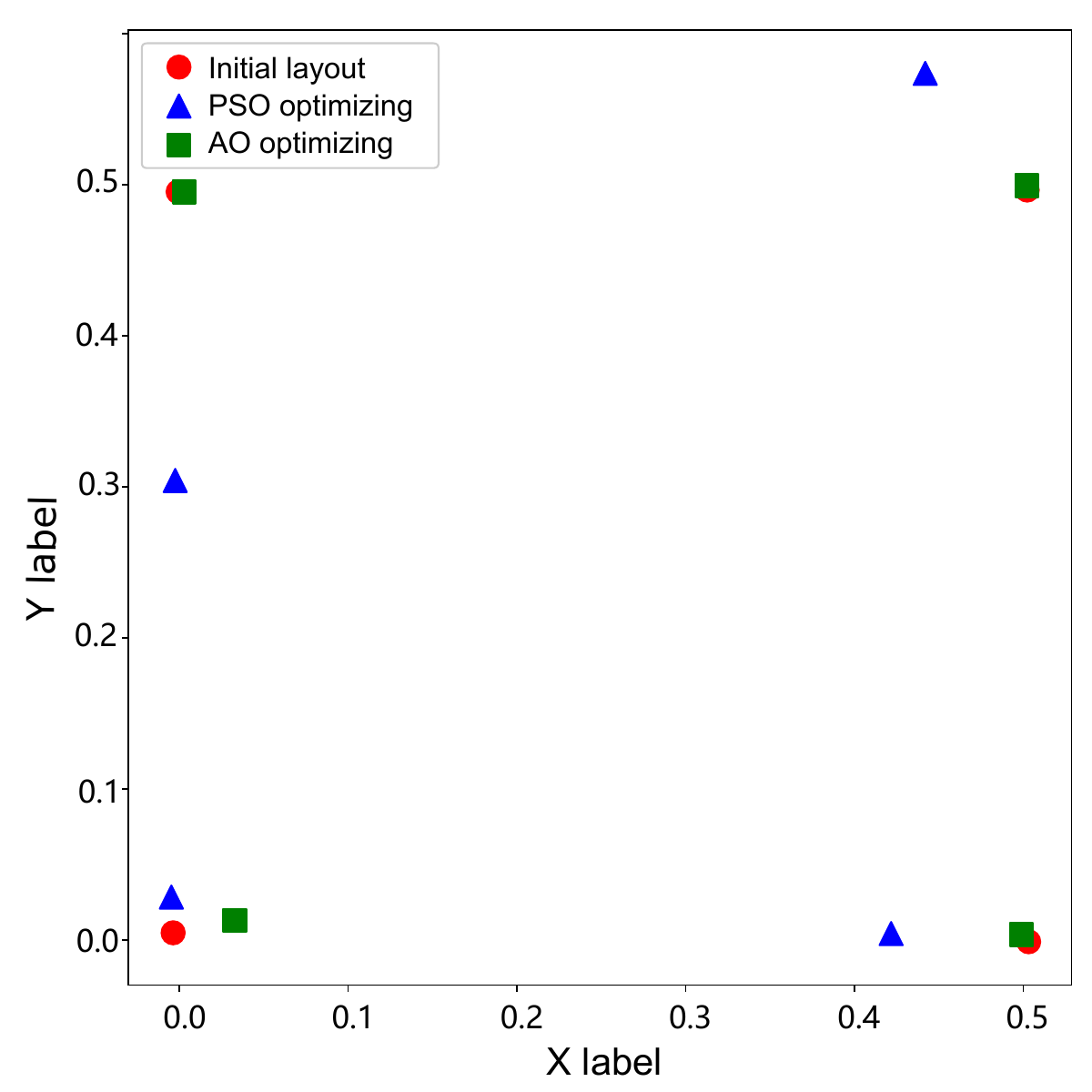} 
	\caption{Antenna layout comparison: initial vs. optimal.}
	\label{fig:layout}
\end{figure}

Figure \ref{fig:layout} shows a comparison between the UPA array and the optimized antenna positions by using FA array. The red circles denote the UPA layout, which are characterized by a regular square structure, the blue triangles represent the antenna positions and precoding matrix optimized via PSO algorithm and the green squares show the result of AO algorithm. The optimal layout demonstrates a clear asymmetry, which indicates that the antenna positions have been adaptively adjusted to enhance channel entropy source. Interestingly, the optimal position for KGR did not deviate much from the initial UPA layout, which suggests that the energy consumption caused by the antenna movement can be small.

\begin{figure}[htbp]
	\centering
	\includegraphics[height=5.5cm]{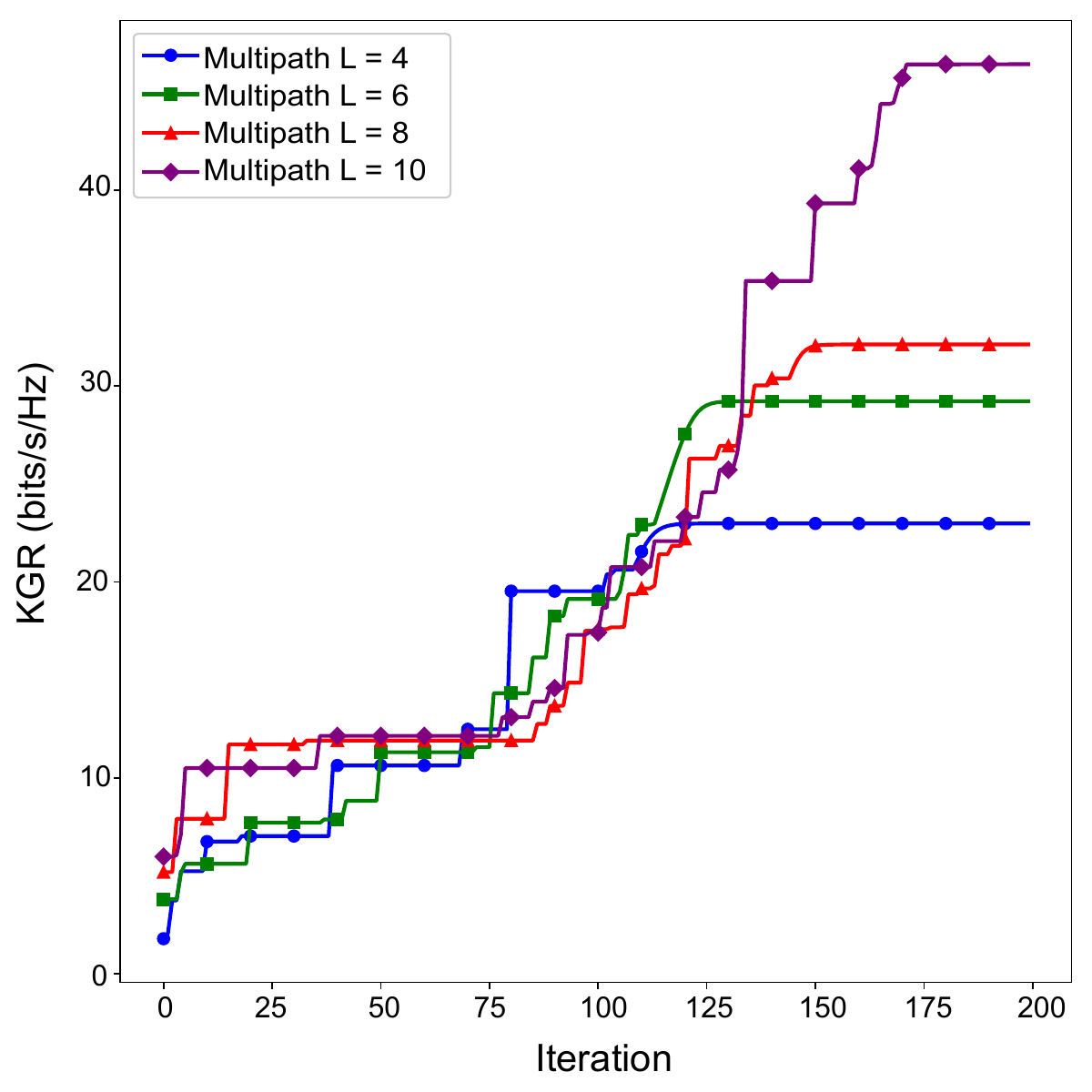} 
	\caption{KGR with different multipath components.}
	\label{fig:multipath}
\end{figure}

Figure \ref{fig:multipath} illustrates the convergence of the KGR under different multipath numbers, i.e, $L = 4, 6, 8, 10$ with the PSO algorithm. It is observed that the PA enabled PLKG is effectiveness in poor multipath environments, and the more diverse the multipath is, the better the KGR is. For example, when $L = 10$, the achieved key rate exceeds 40~bits/s/Hz, while for $L = 4$, the system converges earlier but has a much lower KGR. The reason for different convergence speeds is that the larger number of multipaths increases the search space of the PSO algorithm, which results in a slower convergence. Furthermore, it can be found that the proposed algorithm with fewer number of multipaths shows a relatively gentle improvement in the curve of KGR. When the number of multipaths reaches a certain bound, i.e., $L = 10$, the KGR experiences a significant increase. This result confirms that the proposed algorithm can make better use of the degrees of freedom (DoF) in rich multipath environments.

\section{Conclusion}

This paper studied a novel FA enabled PLKG under the quasi-static wireless environments. We proposed a joint optimization framework based on PSO algorithm that simultaneously adapts the antenna positions and the precoding matrix to maximize the KGR. Moreover, to reduce computational complexity, we developed an AO algorithm combining PGD and PSO algorithm, which separately optimizes the precoding and antenna layout. Simulations shown that since the FA array brings additional DoF, it can significantly enhance the KGR compared with the conventional benchmarks.

\bibliographystyle{IEEEtran}  
\bibliography{references}     

\end{document}